\documentstyle[preprint,aps]{revtex}
\draft
\begin{document}

\title{Mean field effects in heavy-ion collisions at AGS energies}
\bigskip
\author{Bao-An Li, C. M. Ko, and G. Q. Li}
\address{Cyclotron Institute and Physics Department,\\
Texas A\&M University, College Station, TX 77843}
\maketitle

\begin{quote}
The question whether mean field effects exist in heavy-ion collisions
at AGS energies is studied in the framework of A Relativistic
Transport (ART) model. It is found that in central collisions
of Au+Au at $P_{beam}/A=$11.6 GeV/c a simple, Skyrme-type nuclear
mean field satisfying the causality requirement reduces the maximum
baryon and energy densities reached in the cascade model by
about 30\% and 40\%, respectively.
The mean field affects the inclusive, single particle observables
of various hadrons by at most 20\%. We show, however, that
the mean field causes a factor of 2.5 increase in the strength
of the baryon transverse collective flow.

\end{quote}
\newpage
The purpose of relativistic heavy-ion collisions at AGS energies is to
study the properties of hot and dense hadronic matter
and the possible phase transition to the quark-gluon plasma.
A vast body of data have been collected and analysed during the past
years\cite{qmatter}. Comparisons of these data with theoretical
models, such as RQMD \cite{rqmd}, ARC\cite{arc} and QGSM\cite{qgsm},
have revealed much interesting physics. In particular, a picture of
nearly complete stopping of baryons in central heavy-ion collisions
at AGS energies has emerged from these studies.
It has been shown that baryon and energy densities up to 10 times that of
nuclei in their ground state have been reached in these collisions.
This has led to the suggestion that the energy and baryon densities
achieved during central heavy-ion collisions at AGS energies may be
high enough to form the quark-gluon plasma\cite{kapusta}.
However, the above suggestion and many speculations critically
rely on the inferred maximum energy and baryon densities
reached in the reaction.

In this Letter, we report on the main aspects of
a relativistic transport (ART) model for heavy-ion collisions at AGS energies,
and study, in particular, effects of the nuclear mean field
which has been ignored in ARC.
Our motivations of studying the nuclear mean field effects
are mainly the following. Firstly,
we believe that the mean-field potential is not negligible in heavy-ion
collisions at AGS energies. Although the forward scattering amplitudes
of hadron-hadron collisions in the high energy limit have been
found approximately purely imaginary\cite{wong}, the AGS energies
may not be high enough for the real part of the scattering
amplitude to completely vanish. Of course, the form and strength of
the corresponding mean field in the hot and dense hadronic matter is
highly uncertain and has been a subject of much discussions.
Secondly, although the kinetic energy is much higher than the potential
energy in the early stage of the reaction, particles are gradually
slowed down and the mean field plays an increasingly important role
as the reaction goes on. In particular, the repulsive mean field in
the high density region tends to keep particles from coming too close
to each other and therefore reduce the maximum energy and baryon
densities reached in the reaction should there be no mean field.
Moreover, in the expansion phase of the reaction mean-field effects
are expected to be even stronger.
It is therefore necessary to study how the compression and expansion
are affected by the nuclear mean field. Finally, what are the experimentally
observable consequences of the nuclear mean field ?

Our relativistic transport model is developed from the well known
Boltzmann-Uhling-Uehlingbeck (BUU) model (e.g. \cite{bertsch,li91}) for
intermediate energy heavy-ion collisions.
Here we report on the extension of the model to AGS energies
with a reasonable amount of details.
We have included in the model the following
baryons: $N,~\Delta(1232),~N^{*}(1440),~N^{*}(1535),~\Lambda,~\Sigma$,
and mesons: $\pi,~\rho,~\omega,~\eta,~K$, with their explicit isospin degrees
of freedom. Both elastic and inelastic collisions among most of these
particles are simulated as best as we can using as much input from
the experimental hadron-hadron data as possible\cite{data1}.
Most of the inelastic hadron-hadron collisions
are modeled through the formation of resonances.
The finite lifetimes of these resonances effectively take into account
the formation time of newly produced secondaries.
More specifically, we have included in the model
\begin{eqnarray}
&&NN\leftrightarrow N\Delta,~NN^{*}(1440),~NN^{*}(1535),\\
&&NN\leftrightarrow \Delta\Delta,~\Delta N^{*}(1440),\\
&&NN\rightarrow NN\rho,~NN\omega,~\Delta\Delta\pi,\\
&&NN\rightarrow \Delta\Delta\rho,\\
&&N\Delta\leftrightarrow NN^{*}(1440),~NN^{*}(1535),\\
&&\Delta\Delta\leftrightarrow NN^{*}(1440),~NN^{*}(1535),\\
&&\Delta N^{*}(1440)\leftrightarrow NN^{*}(1535),
\end{eqnarray}
and those producing kaons
\begin{eqnarray}
&&NN\rightarrow N\Lambda(\Sigma)K,~\Delta\Lambda(\Sigma)K,\\
&&NR\rightarrow N\Lambda(\Sigma)K,~\Delta\Lambda(\Sigma)K,\\
&&RR\rightarrow N\Lambda(\Sigma)K,~\Delta\Lambda(\Sigma)K,
\end{eqnarray}
where R is $\Delta,~N^{*}(1440)$ or $N^{*}(1535)$.
Cross sections parameterized by VerWest et al.\cite{verwest} are
used for the single $\Delta$ and $N^{*}(1440)$ production in
processes (1).
The cross section of $N^{*}(1535)$ production is estimated
from that of $\eta$ production\cite{wolf}.
Cross sections for the double resonances production in processes (2) are taken
to be the same for all possible isospin channels,
this is supported by recent calculations based on the one-boson-exchange
model\cite{aichelin}. Numerically, it is estimated
by subtracting from the inclusive $2\pi$ production cross
section\cite{data1} the contribution of $NN\rightarrow NN\rho$
and the $2\pi$ decay of the $N^{*}(1440)$ in the
$NN\rightarrow NN^{*}(1440)$ process.
The cross sections for $\rho$ and $\omega$ production in channels
(3) are taken directly from the experimental data\cite{data1}.
The cross section for the process $NN\rightarrow \Delta\Delta\pi$
is taken as the difference between the inclusive $3\pi$ and the $\omega$
production cross section. We attribute the difference between
the experimental total nucleon-nucleon inelastic cross
section and the sum of cross sections for channels (1) to (3)
as well as the kaon production
cross sections of channels (8) to the process
$NN\rightarrow\Delta\Delta\rho$. This is done to ensure the total
inelasticity of nucleon-nucleon collisions as only a limited, though large,
number of reaction channels have been incorporated so far.
The bias introduced by this approximation towards the cross section
of the quasi-$4\pi$ production process $NN\rightarrow \Delta\Delta\rho$
is very small.
Cross sections for channels (5) to (7), (9) and (10) are taken
to be the same as in nucleon-nucleon collisions having the same center of mass
energy and total charge.
Cross sections for the inverse
processes are calculated by using the detailed balance.
Masses of baryon and meson resonances are generated according to
the single or joint Breit-Wigner distributions with momentum dependent
widths using the rejection method.

One can also separate meson-baryon collisions into elastic and inelastic
parts. We model elastic collisions through both the formation of
baryon resonances, i.e., $
\pi N\leftrightarrow \Delta (N^{*}(1440), N^{*}(1535)) $ and
$\eta N\leftrightarrow N^{*}(1535)$,
as well as direct processes, i.e.,
$\pi (\rho)+N(\Delta, N^{*})\rightarrow \pi (\rho)+N(\Delta,N^{*})$,
and $K+N\rightarrow K+N$. The formation of the three baryon resonances accounts
almost completely the $\pi+N$ elastic cross sections at low energies. At higher
energies, for $\pi^{-}+p$ at $\sqrt{s}\geq 1.7 $GeV, for example,
the elastic cross section is about 7 mb and is mainly due to the formation
of higher resonances which are not included in the present model.
We therefore attribute the difference between the experimental elastic
cross section and the contribution from the three baryon resonances
to the direct process $\pi+N\rightarrow \pi+N$. For experimentally unknown
cross sections, such as, $\pi^{0}+N$, $\pi+\Delta (N^{*})$ and
$\rho+N$, we calculate them using the resonance model. Neglecting the
interference between resonances, one has
\begin{equation}
\sigma(M+B)=1.3\frac{\pi}{k^{2}}\sum_{R}\frac{(2J_R+1)}{(2S_M+1)(2S_B+1)}
\frac{\Gamma^2_R(M+B)}{(\sqrt{s}-m_R)^2
+0.25\Gamma_{R}^2(total)}.
\end{equation}
The factor 1.3 is obtained by fitting to the
high energy part of the $\pi^{+}+p$ data. For calculating
the cross sections of direct channels, R runs over all baryon resonances
with masses up to 2 GeV.

The $\pi+N$ inelastic collision mainly goes through the production of
pions and kaons. Similar to the treatment of baryon-baryon collisions,
we model the inelastic $\pi+N$ collisions through the production of
resonances, namely,
$\pi+N\leftrightarrow \Delta+\pi(\rho,\omega)$ and
$\pi(\rho,\omega)+N(\Delta, N^{*})\rightarrow \Lambda (\Sigma)+K$.
All experimental cross sections for $\pi+N$ collisions with final states
having two and three pions are attributed to the production of
$\Delta\pi$ and $\Delta\rho$, respectively. The difference between the
experimental total $\pi+N$ inelastic cross section and cross sections for
the production of two and three pions as well as kaons are attributed to
the production of $\Delta\omega$. Cross sections for the production of
kaons in $\pi (\rho,\omega)+N(\Delta, N^{*})$ collisions are
taken to be the same as that in the $\pi+N$ collision.

We model pion-pion elastic collisions through the formation of
the $\rho$ meson and the direct process.
The latter takes into account the case when the quantum numbers
of colliding pions forbid the formation of the $\rho$ meson.
These cross sections are taken mainly from ref.\ \cite{bert}.
The inelastic collisions among mesons are modeled through
the production of $K\bar{K}$. The cross section for this process is
highly uncertain, here we use the cross section calculated from the
$K^{*}$-exchange model of ref.\ \cite{ko}.

The ``leading'' particle behaviour in energetic collisions is
ensured by requiring the outgoing baryons to have the same or similar
quantum numbers as the incident ones so that their longitudinal
directions are retained when performing the momentum
transformation from the baryon-baryon
c.m. frame to the nucleus-nucleus c.m. frame. The longitudinal and
transverse momenta of baryons in the final state is assigned
according to the following distribution under the constraint of total
energy and momentum conservations\cite{data2}
\begin{equation}
\frac{d^{2}N}{d^3\vec{p^*}}\propto (1.+0.5x^{*}-0.9x^{*^2})
(e^{-4p_{t}^{2}}+0.5e^{-10p_{t}}),
\end{equation}
where $x^{*}$ is the scaled longitudinal momentum in the c.m. frame, i.e.,
$x^{*}=2p^*_z/\sqrt{s}$. The distribution has the properties of the naive
scaling, $p_t$ and $p^{*}_{z}$ factorization and the soft, energy
independent transverse momentum distribution.
With this distribution and the cross sections discussed above, we can
well reproduce the longitudinal and transverse momentum distributions of
protons and pions in $pp$ collisions at $p_{beam}=$12 GeV/c and below.
The test against $pp$ data together with many other details of the model
will be published separately\cite{li95}.

In the framework of the ART model outlined above, we shall concentrate on
studying the effects of nuclear mean field. Without much reliable knowledge
about the nuclear equation of state in hot and dense medium, we
use here the simple, Skyrme-type parameterization widely used at Bevalac
energies and below,
\begin{equation}\label{field}
U(\rho)=-358.1(\frac{\rho}{\rho_{0}})+304.8(\frac{\rho}{\rho_{0}})^{1.167}.
\end{equation}
This is the so-called soft equation of state which satisfies the
causality requirement upto about $7\rho_0$. The standard test-particle method
is used to calculate the global baryon density $\rho_g$ and energy density
$e_g$ in the nucleus-nucleus c.m. frame on a lattice with the
cell size of 1 {\rm fm}$^3$. The local baryon density and energy density
in each cell are obtained by
$
\rho_{l}=\rho_{g}/\gamma
$
and
$
 e_{l}=e_{g}/\gamma,
$
where $\gamma$ is the Lorentz factor in each cell.

Let us first study effects of the mean filed on the
creation of high baryon and energy densities.
We show in Fig.\ 1 the evolution of the
local baryon and energy densities in the central cell during
the reaction of Au+Au at $P_{beam}/A$=11.6 GeV/c and b=0 for
three different cases. For the cascade case, it is seen
that the maximum baryon density of about $9\rho_0$ and the maximum
energy density of about 3.6 {\rm GeV/fm}$^{3}$ are reached at
about 4 {\rm fm}/c. The high energy density matter
lasts for about 5 {\rm fm}/c. A more detailed examination of the
density contour plots shows that the high energy density region
with $e_{l}\geq$ 2.0 {\rm GeV/fm}$^{3}$ has the maximum volume
of about 200 {\rm fm}$^3$ at about 4 {\rm fm}/c.
It is worth mentioning that the main features of the cascade model
calculations are very similar to that of ARC.

The currently estimated critical baryon and
energy densities for forming the quark gluon plasma is about
$5\rho_0$ and 2.5 {\rm GeV/fm}$^3$, respectively\cite{wong}.
According to the above cascade model prediction, it is highly
possible to form the quark-gluon plasma at AGS energies.
However, the lifetime and volume of the high density region are
significantly reduced by the repulsive mean field.
Noticeably, the maximum baryon and energy densities are
reduced to about $7\rho_0$ and 2.6 {\rm GeV/fm}$^{3}$, respectively,
by using the soft nuclear equation of state.
With a stiff equation of state corresponding to the compressibility
of $K=380$ {\rm MeV} the reduction is even larger although
the mean field has almost no effect in the early
stage of the reaction when the kinetic
energy is much higher than the potential energy.
Since the stiff equation of state violates causality already at
about $3\rho_0$, we will only use the soft equation of state in the following.
The reduction of the maximum baryon and energy densities due to the
mean field may be large enough to affect significantly
the kind of physical processes that can happen during the reaction.

The validity of any hadronic model for heavy ion collisions at AGS energies
is naturally limited by the possible phase transition to the quark-gluon
plasma.
With this precaution in mind, one can use predictions of hadronic models
as a baseline in searching for new phenomena.
The formation of the quark-gluon plasma or a mixed phase of hadrons and
quark-gluon plasma is expected to reduce the
pressure of the system and leads to a softened nuclear
equation of state.
It is therefore interesting to search for
experimental observables that are sensitive to
the nuclear equation of state. A careful study on the transverse
momentum and rapidity distributions of protons, pions
and kaons for the reaction of Au+Au at $P_{beam}/A=$11.6 GeV/c has
been done by comparing calculations
with and without the mean field. Much to our surprise, both calculations
can well reproduce the inclusive, single particle data within current
systematic and statistical error bars of the data.
The largest difference of about 20\% is found for the proton transverse
momentum distribution with a slight shift to higher transverse momenta in
the case using the soft equation of state. This difference is, however,
compatible with the current systematic error bars of the data.
These results will be published
elsewhere\cite{li95}.

Hinted by the findings at Bevelac energies
that the collective variables or correlation functions, unlike the single
particle observables, are very sensitive to the nuclear equation of state,
we now turn to the analysis of the baryon transverse collective
flow\cite{dani}.
Here we use the standard flow analysis as in ref.\ \cite{bertsch}.
It was shown recently in ref.\ \cite{gale} that
an improved flow analysis with the explicit conservation of
reaction plane and angular momentum in individual hadron-hadron
collisions results in an increase of the transverse momentum by about
8\% to 23\% at Bevalac energies. This small enhancement will not affect
our discussions and conclusions in the following.

We show in Fig.\ 2 the average transverse momentum of nucleons in the reaction
plane as a function of rapidity for the reaction of Au+Au at
$P_{beam}/A=$11.6 GeV/c and impact parameters of 2, 6 and 10 {\rm fm}.
Significant differences exist between
calculations with and without the mean field for the reaction at
all of the three impact parameters. In particular, the flow
parameter defined as the slope of the transverse momentum
distribution at midrapidity is about a factor of 2.5 larger
in the case with the mean field. The strength of
the so-called ``bounce-off'' effect at target or projectile
rapidities is also much stronger in calculations with the mean field.
It is also seen that the collective flow is the strongest in the midcentral
collisions.

To see why the collective flow is a sensitive probe of the
dynamics in the high density region, we show in Fig.\ 3 the
flow parameter as a function of time for the Au+Au reaction
at impact parameters of 2 {\rm fm} and {\rm 6 fm}. It is clear that
the flow is mainly generated in the high density region and saturates
in the expansion phase. The ratio of final flow parameters
in the calculations with and without the mean field is
about 2.5 in the reactions at both impact parameters.
This difference is large enough for a clear distinction between
models when the experimental data is available.
For the reaction at an impact parameter of 6 fm, the flow parameter
undergoes a reduction before saturation. This is due to the reflection of
hot baryons from the cold spectator nucleons.
It is interesting to note that the decrease of pressure in the
high density region due to the possible phase transition to guark-gluon
plasma is expected to reduce the strength of collective
flow. An experimentally measured, significantly smaller flow
parameter than the cascade prediction will be a strong indication
of the quark gluon plasma formation at AGS energies.

In summary, we have developed a new relativistic transport model for heavy-ion
collisions at AGS energies. Within the framework of this model, we have found
that the mean field can significantly affect the maximum energy and baryon
densities reached in the reaction. The transverse collective flow analysis is
suggested as a useful tool to disentangle theoretical models and
a possible indicator of the formation of quark-gluon plasma at AGS energies.
\medskip

We would like to thank many participants to the program
``Hot and dense nuclear matter'' at the National Institute for Nuclear
Theory at Seattle for interesting discussions. In particular, we
are indebted to W. Bauer, L.P. Csernai, P. Danielewicz,
S. Das Gupta, V. Koch, J. Randrup and Gy. Wolf for their
constructive suggestions and criticisms.
This research was supported in part by the NSF Grant No. PHY-9212209
and the Welch Foundation Grant No. A-1110.

\section*{Figure Captions}
\begin{description}
\item{\bf Fig.\ 1}\ \ \
Evolution of the central local baryon density and
energy density in the reaction of
Au+Au at $P_{beam}/A=$11.6 GeV/c and b=0.
\item{\bf Fig.\ 2}\ \ \
Baryon average transverse velocity in the reaction plane as a function of
rapidity for Au+Au reactions at $P_{beam}/A$=11.6 GeV/c and three different
impact parameters.
\item{\bf Fig.\ 3}\ \ \
Time evolution of the flow parameter in the reaction of
Au+Au at $P_{beam}/A$=11.6 GeV/c and the impact parameter
of 2 and 6 fm.

\end{description}

\end{document}